\documentstyle[12pt]{article}
\newcommand{\be}{\begin{equation}}
\newcommand{\ee}{\end{equation}}
\newcommand{\bea}{\begin{eqnarray}}
\newcommand{\eea}{\end{eqnarray}}
\def\pd{\partial}
\begin{document}

\begin{titlepage}

\title{Infinite series solutions of the symmetry equation for the $1 +2$
dimensional continuous Toda chain}

\author{D.B. Fairlie\\
\smallskip\\
{\it Department of Mathematical Sciences}\\
{\it University of Durham, Durham DH1 3LE, England}\\
\medskip\\
and A.N. Leznov\thanks{on leave from {\it  Institute for High
Energy Physics, 142284 Protvino, Moscow Region, Russia}}\\
\smallskip\\
 {\it Max-Planck-Institut f\"ur Mathematik}\\
{\it Gottfried-Claren-Strasse 26,}\\
{\it 53225 Bonn, Germany.}}

\maketitle

\begin{abstract}  A sequence of solutions to the equation of symmetry for the
continuous Toda chain in $1+2$ dimensions is represented in explicit form.
This fact leads to the supposition that this equation is completely integrable.
\end{abstract}

\end{titlepage}
\section{Introduction}
The problem of the continuous Toda chain arose as a rediscovery of the earlier
results of Darboux \cite{darboux} and Fermi, Pasta and Ulam
\cite{fermi}. In a wider context, this problem illustrates features common to
all other
infinite dimensional integrable chains. The problem consists of taking the
limit of the interparticle distance in the discrete Toda chain in two
dimensions to zero, giving the continuous version described by the
equation\cite{boy}\cite{park}\cite{sav};
\begin{equation}
\frac{\pd^2}{\pd x\pd y}\log{U}=\frac{\pd^2}{\pd z^2}U.
\label{cont}
\ee

The discrete Toda chain in two dimensions takes the form of the infinite set of
equations
\be
\frac{\pd^2}{\pd x\pd y}\log U_n=U_{n+1}-2U_n+U_{n-1}.
\label{disc}
\ee
With a suitable rescaling, the limit of equation (\ref{disc}) is (\ref{cont}).
Under appropriate boundary conditions the equations of the infinite dimensional
Toda chain (\ref{disc})  reduce to a finite-dimensional (classical dynamical
system).
Such a Toda chain with fixed ends $(U_0 =0,\ U_n=0)$ is an exactly integrable
system.
Its general solution was found in \cite{L1}. In the case of periodic boundary
conditions $(U_i=U_{i+N})$ the resulting system  coincides with affine Toda
chain
connected with Kac-Moody algebras $A_N$. In this case it is completely
integrable and it is possible to find in explicit form  only c-number
parametrical soliton-like solutions \cite{L2}. In the one dimensional case,
where $U$ depends upon $(x,\ y,\ z)$ only through the combinations $(x+y,\ z)$
the general implicit solution has been found by R.S. Ward \cite{ward}.

The obvious question arises; to what class of equations does equation
(\ref{cont})
belong? To obtain a solution of this equation by means of a limiting procedure
on a solution of (\ref{disc}) does not seem feasible as solutions of this set
of
equations appear in the form of ratios of determinants. The limiting procedure
is equivalent to finding the limits of infinite determinants, which is always
rather difficult.
For the investigation of equation (\ref{cont}), we shall use  classical
group theoretical methods \cite{Ov}\cite{Ol},
 as a technique, by constructing the solutions of the corresponding symmetry
equation , which takes the form
\begin{equation}
\frac{\pd T}{\pd x}=U\int\frac{\pd^2 T}{\pd z^2}dy.
\label{symm}
\ee
(The term `symmetry equation' is the one most often used in the literature to
designate the equation describing the variation of the dependent variables
with respect to a parameter of the solution.)
In spite of the known example of the solutions of the symmetry equation for the
analogous Sine Gordon case \cite{Sh1},\cite{Sh2} which are expressed in terms
of multiple derivatives, the solutions of (\ref{symm}) which we shall obtain
here are represented as linear combinations of repeated integrals, which are
constructed
using precise algorithmic rules.
\section{Solution of the symmetry equation}
The symmetry equation for our problem  can be derived either indirectly, by
taking it as the limit of the symmetry equation for the discrete chain,
or else directly from (\ref{cont}), by differentiating (\ref{cont}) with
respect to an arbitrary parameter $\tau$ and setting $\displaystyle{S =
\frac{\pd U}{\pd\tau}}$. Upon integration over $y$, this yields the equation
\begin{equation}
\frac{\pd}{\pd x}\left(\frac{S}{U}\right)=\int\frac{\pd^2 S}{\pd z^2} dy.
\label{start}
\ee
 In this equation change variables to $T$ defined by
\be
 S\to \frac {\pd T}{\pd x}
\label{subs}
\ee
Then the equation becomes (\ref{symm}) , after integration by $x$ and
differentiation once by $y$. (Another formulation of the symmetry equation
consists in making the  substitution $\displaystyle{S\to U\frac {\pd W}{\pd
z}}$.
This yields the equation $ W_x=\int dy (UW_z)_z $ which may be solved by
similar methods)

The derivation of the symmetry equation for the discrete chain will be
discussed in the next  section.
The method of approach is based upon the solution of a discrete version of
equation (\ref{symm}) derived in \cite{varat}.
First of all we may remark that $ T = T^0 =U$ is a solution of (\ref{symm}).
Let us seek a solution of (\ref{symm}) in the form
\be
T= T^0\alpha^0.
\ee
The equation obeyed by $\alpha^0$ following from (\ref{symm}) (taking into
account (\ref{cont})) takes the form
\be
\alpha^0_x+\alpha^0\int T^0_{zz}dy=\int (T^0\alpha^0)_{zz}dy,
\label{start1}
\ee
Let us represent $\alpha^0$ as an integral $\alpha^0\mapsto \int \alpha^0dy$,
retaining the same symbol for the integrand and differentiate (\ref{start1})
with respect to the argument $y$. We  obtain
\be
\alpha^0_x+\alpha^0\int T^0_{zz}dy=2T^0_z\int\alpha^0_{z}dy+T^0\int
\alpha^0_{zz}dy.
\label{start2}
\ee
Let us attempt to find a solution of (\ref{start2}) with the aid of the ansatz
\be
\alpha^0=T^0_z\alpha^1_{z}+T^0 \beta^1.
\label{ansatz}
\ee
This representation is suggested by the fact that the right hand side of
(\ref{start2}) contains terms with factors
$T^0_z,\ T^0$. After substitution of this ansatz and setting the coefficients
 of  $T^0_z$ and $T^0$ to zero in the resulting expression (this is an
addditional
assumption), the pair of equations which the functions $\alpha^1$ and $\beta^1$
satisfy will be
\bea\label{pair}
\alpha^1_x+2\alpha^1\int T^0_{zz}dy&=& 2\int \
\alpha^0_z dy,\nonumber\\
\beta^1_x+2\beta^1\int T^0_{zz}dy+\alpha^1\int T^0_{zzz}dy&=&\int
\alpha^0_{zz}dy.
\eea
If the derivative of the first equation with respect to the variable $z$ is
subtracted from twice the second we  obtain as a consequence the equation
\be
(2\beta^1-\alpha^1_z)_x+2
(2\beta^1-\alpha^1_z)\int T^0_{zz}dy=0.
\label{constraint}
\ee
{}From the last equation we conclude that among the solutions of
(\ref{constraint}) are those for which
\be
\beta^1=\frac{1}{2}\alpha^1_z.
\label{constraint2}
\ee
Resubstitution of this solution for $\beta^1$ into the first of (\ref{pair})
gives the following equation to determine $\alpha^1$;
\be
\alpha^1_x+2\alpha^1\int T^0_{zz}dy=\int(2\alpha^1T^0_{zz}+3\alpha^1_zT^0_z+
\alpha^1_{zz}T^0)dy.
\label{start3}
\ee
If $\alpha^1$ is represented in integral form as $\alpha^1\mapsto \int
\alpha^1dy$,
the  form of (\ref{start3}) after differentiation with respect to $y$
becomes
\be
\alpha^1_x+2\alpha^1\int
T^0_{zz}dy=3T^0_z\int\alpha^1_zdy+T^0\int\alpha^1_{zz}dy.
\label{start4}
\ee
This equation differs from the equation (\ref{start2}) for the determination
of $\alpha^0$ only by numerical factors in the various terms.
Repeating the above trick for $\alpha^1$, $\alpha^2$ etc. we finally arrive at
the  sequence
\be
\alpha^{k-1}=T^0_z\alpha^k_{z}+T^0 \beta^k,
\label{ansatzk}
\ee
where $\displaystyle{\beta^k=\frac{1}{k}\alpha^k_z}$ and $\alpha^k$ has the
form
\be
\alpha^k_x+(k+1)\alpha^k\int
T^0_{zz}dy=\int((k+1)\alpha^kT^0_{zz}+(k+2)\alpha^k_zT^0_z+\alpha^k_{zz}T^0)dy.
\label{start5}
\ee
This last equation possesses the obvious solution
\be
\alpha^k =1.
\label{soln}
\ee
The final result may be represented in two forms; one of an algorithmic nature,
the second in the form of repeated integrals and derivatives. In the first,
the structure of a particular solution to the symmetry equation takes the
form
\be
T^n =T^0\prod^n_{j=1}((j+1)D_j+\sum^n_{k=j+1}D_k)\overbrace{\int T^0dy_1\int
T^0dy_2\cdots\int T^0dy_n}^n,
\label{algol}
\ee
where $D_j$ denotes the operation of  differentiation with respect to $z$ of
the integrand $T^0$ situated at the $j$ th place in the $n$ repeated integrals
in (\ref{algol}).
An alternative expression for this solution is given by the formula (setting
$T^0=t$);
\be
T^n = t\int\frac{dy_1}{t}\frac{\pd}{\pd z}(\frac{t^2}{2}\int \frac{dy_2}{t^2}
\cdots\frac{\pd}{\pd z}(\frac{t^j}{j}\int \frac{dy_j}{t^j}\cdots \int\frac{\pd
t}{\pd z}dy_n))\cdots).
\ee
(note that $t^j$ denotes the $j$ th power of $T^0$, not $T^j$ and there are $n$
iterated integrals in the above expression).
\section{Symmetry equation for the discrete Toda chain}
In this section we give the corresponding analysis for the case of the
discrete Toda chain.
The symmetry equation for the discrete Toda equation is obtainable from a
generalisation of the Darboux transformation which takes the form, for two
independent functions
\bea
(\log U_n)_{xy}&=&V_n({U}_{n+1}-U_n)-{V}_{n-1}(U-{U}_{n-1}),\nonumber\\
(\log V_n)_{xy}&=&{U}_{n+1}({V}_{n+1}-V)-U (V-{V}_{n-1}),
\label{gen}
\eea
 where  (\ref{gen}) is viewed
as a set of recurrence relations connecting functions evaluated at the nodes of
a chain. More generally $\displaystyle{U}_{n+k}$ can be viewed as either  the
value of the function $U_n$ at a point $k$ places to the right, or else the $k$
th iterate of the function $U_n$.

If  $V=1$,
this reproduces the standard discrete Toda chain (\ref{disc}). Also
$\displaystyle{f={U}_n{V}_{n+1}}$ and $g=U_nV_n$ also satisfy the same discrete
equation
and in the continuum limit $f\to g$ the continuous Toda chain (\ref{cont}). The
symmetry equation
\be
(T_n)_x=U \int dy [V_n({T}_{n+1}-T_n)-{V}_{n-1}(T_n-{T}_{n-1})]
\label{dissymm}
\ee
may be solved by analogous methods to those in the previous section. (This
equation is equivalent to equation (2.16) of \cite {varat})

First of all, it is clear that $T_n=U_n$ is a solution. Then set
$T_n=U_n\int\alpha^0_n dy$. The equation then becomes
\bea
&&(\alpha^0_n)_x+\alpha^0_n\int
[V({U}_{n+1}-U_n)-{V}_{n-1}(U_n-{U}_{n-1})]dy=\\
&&V_n{U}_{n+1}\int({\alpha^0}_{n+1}-\alpha^0_n)dy-{V}_{n-1}{U}_{n-1}\int({\alpha^0_n}-{\alpha^0}_{n-1})dy\nonumber
\label{rec}
\eea
This equation may be rewritten in the form
\bea
&&\alpha^0_x+\alpha^0\int [{f}_{n+1}-f_n+{g}_{n-2}-g_n)]dy\nonumber\\
&&={f}_{n+1}\int({\alpha^0}_{n+1}-\alpha^0_n)dy-{g}{n-1}\int({\alpha^0_n}-{\alpha^0}_{n-1})dy
\label{rec2}
\eea
which is equation (3.1) of \cite{varat}. The discussion proceeds as in that
article;   $\alpha^0_n$  is represented in terms of the coefficients of the
integrals of $f,g$ and two new unknown functions $\alpha^1_n,\ \beta^1_n$ as
\be
\alpha^0_n= \alpha^1_n\int{f}_{n+1}dy+ \beta^1_n
\int{g}_{n-1}dy\label{ans2}
\ee
and a pair of  equations  similar to (\ref{rec2}) for $\alpha^1_n,\ \beta^1_n$
obtained. It turns out that  $\beta^1_n=-{\alpha^1}_{n-1}$ is a sufficient
condition
for the consistency of this pair, and the remaining equation for $\alpha^1_1$
takes a similar form to (\ref{rec2}). The iteration proceeds as before, and in
this manner
a class of solutions  can be obtained in the form
\bea\label{24}
T^N_n&=&T^0_n \prod_{i=1}^N (1-L_i\exp[-(i+1) d_i-\sum_{k=i+1}^N
d_k])\nonumber\\
&\times&\int dy_1  {f}_{n+1} \int dy_2  {f}_{n+2}......\int dy_N {f}_{n+N}
\eea
where the symbol $\exp d_s$ means that the argument of the s-th term of
repeated integral ($....\int dy  {f}_{n+h}...\to ...\int dy  {f}_{h+n+1}...$)
in (\ref{24})  should be shifted by unity and the symbol $L_p$ means the
exchange of $ {f}_{n+r}$ and $ {g}_{n+r}$ in the corresponding p-th term
$...\int dy {f}_{n+r}...\to...\int dy {g}_{n+r}...$.
The solution of the previous section can be recovered by setting $f_n=g_n=U$.
\section{General solution in the $1+1$ dimensional case}
In this section we demonstrate the connection between the exact integrability,
in the sense of the existence of a general implicit solution of the $(1+1)$
dimensional continuous Toda chain and the complete solution of its symmetry
equation
It is well known that in the `one dimensional' case,
($\displaystyle{\frac{\pd}{\pd x}=\frac{\pd}{\pd y}}$) equation (\ref{cont}) is
exactly integrable. Its solution has been obtained in \cite{ward}. We now
want to show that in this case the general solution of the symmetry equation
can be obtained. The result will be quoted leaving the reader to check its
validity by direct computation.

The continuous Toda system in this case may be written in terms of two
functions $u,\ v$ which obey the equations
\be
u_t=uv_x,\ \ v_t=vu_x.
\label{two}
\ee
The corresponding symmetry equations for the functions $U,\ V$ are
\be
U_t=Uv_x+uV_x,\ \ V_t=Vu_x+vU_x.
\label{twos}
\ee
Let $T,X$ be an arbitrary solution of the system
\be
u\frac{\pd T}{\pd u}=\frac{\pd X}{\pd v},\ \ v\frac{\pd T}{\pd v}=\frac{\pd
X}{\pd u}
\label{three}
\ee
The complete solution of this equation has been recorded in \cite{strachan},
where $T$ and $X$ are the conserved charge and current densities respectively
for the system (\ref{two})
Then the solution of the symmetry equation may be represented as
\be
U=Tu_t+Xu_x,\ V=Tv_t+Vv_x
\label{four}
\ee
The general solution of (\ref{two})\cite{strachan} depends on two arbitrary
functions of one argument and the solution of the symmetry equation  also
depends upon two arbitrary functions so (\ref{four}) is indeed the general
solution. Thus existence of the exact integrability of the symmetry equation
and that of the continuous Toda chain in the one dimensional case is
demonstrated.

\section{Conclusion}
The analysis of a sequence of solutions to the symmetry equations of the
discrete Toda chain found in \cite{varat} has been successfully adapted to the
case of the continuous Toda chain. In this case , however we have found that
the solution can be encapsulated into a single iterative formula. The solutions
obtained suggest the possibility that the continuous Toda chain is a completely
integrable system. If this is so, then it will possess multisoliton solutions,
which may be found in explicit form. This is in agreement with the general
ideas put forward in \cite{leznov}. We have also demonstrated that in the case
of the one dimensional chain, where exact integrability holds, the existence of
a complete solution to the symmetry equation.
\section{Acknowledgements}
One of the authors, A.N Leznov wishes to acknowledge the hospitality of the
Max-Planck Institute and is indebted to
  the Grant N RMM000 of the International Scienntific Foundation for partial
support. The other  D. B. Fairlie is grateful to the E.C Human Capital
and Mobility Grant ERB-CHR-XCT 920069 for travelling expenses.

\end{document}